\documentstyle[preprint,aps]{revtex}
\newcommand{\mb}[1]{ {\mbox{\boldmath{$#1$}}}  }

\begin{document}
\draft

\title{Self-consistent interface properties of $d$ and $s$-wave 
superconductors.} 

\author{A.M. Martin and  J.F. Annett}
\address{ University of Bristol, H.H Wills Physics Laboratory, 
Royal Fort, Tyndall Ave,
Bristol BS8 1TL, United Kingdom.}

\date{\today}
\maketitle

\begin{abstract}

We develop a method to solve the Bogoliubov de Gennes equation for
superconductors self-consistently, using the recursion method.  The
method allows the pairing interaction to be either local or non-local
corresponding to $s$ and $d$-wave superconductivity, respectively.
Using this method we examine the properties of various $S-N$ and $S-S$
interfaces.  In particular we calculate the spatially varying density
of states and order parameter for the following geometries (i) $s$-wave
superconductor to normal metal, (ii) $d$-wave superconductor to normal
metal, (iii) $d$-wave superconductor to $s$-wave superconductor.  We show
that the density of states at the interface has a complex structure
including the effects of normal surface Friedel oscillations, the
spatially varying gap and Andeev states within the gap, and the subtle
effects associated with the interplay of the gap and the normal van
Hove peaks in the density of states. In the case of bulk $d$-wave
superconductors the surface leads to mixing of different order
parameter symmetries near the interface and substantial local filling
in of the gap.
\end{abstract}

\narrowtext
\section{Introduction.}
\label{introduction}
\setcounter{footnote}{1}

Interfaces in superconductors, especially high temperature
superconductors, are of considerable interest for both fundamental
physics and for applications.  Many experiments have used both
single electron tunnelling and Josephson effects as probes of the
energy gap and order parameter symmetry in the cuprates.
For example Wollman {\it et al.}\cite{wollman} 
and Sun {\it et al.}\cite{sun} constructed SQUID devices
consisting of junctions between YBa$_2$Cu$_3$O$_7$ (YBCO) and Pb, while
Tsuei {\it et al.}\cite{tsuei} constructed superconducting rings consisting
of YBCO thin films with two or three grain boundary junctions.  
Theoretical analysis of experiments such as these relies to
a large extent on macroscopic symmetry arguments and not on the microscopic
details of the actual interfaces.  However in some cases the microscopic
physics at the interface can be an important factor in understanding
the experimental results. For example, if mixing of different
order parameter symmetries occurs at the interface (because the
interface breaks the bulk tetragonal or orthorhombic symmetry)
the extent of such mixing can only be determined from
microscopic calculations. Similarly, suppression of the order parameter
(either $d$-wave or $s$-wave) near an $S-N$ interface can lead to
significant local density of states within the bulk 
energy gap, and this can complicate the analysis of
single electron tunnelling spectra\cite{kitizawa}. 
The microscopic physics of surfaces and interfaces of high
T$_c$ superconductors are especially interesting because of
the short coherence length, and the probable $d$-wave gap function.

In the past few years there have been a number of
microscopic calculations of surfaces and interfaces in
superconducting systems with a $d$-wave
order parameter. Most of the theoretical results have been obtained 
using tunnelling theory, or Andreev's 
approximation\cite{CRHu,XMT,TK,ZWT,YTanaka,TWZ,TK1} in which the tunnelling
barrier and the order parameter are not found self-consistently.
For tunnel junctions these approximations may be adequate, but we show
below that self-consistency has significant effects for interfaces
with direct contact between the constituents. 
Self-consistent properties of interfaces have previously been computed
using the 
Eilenberger equations \cite{BGZ,MS1,MS2,MS3,BSB,MKM}, which are an 
approximation to
the Bogoliubov de Gennes equation. These self-consistent solutions to
the Eilenberger equations have shown some interesting effects which
have only arisen when the order parameter is calculated in a
self-consistent manner.

In this paper we aim to show how to calculate self-consistent
properties of superconducting interfaces by directly
solving the Bogoliubov de Gennes equations.  This approach has the advantage
that self-consistency can be fully incorporated and also there are
no need for the further approximations of the Eilenberger method.
Our approach makes use of the recursion method\cite{RHaydock}
to solve the Bogoliubov
de Gennes equation on an arbitrary tight binding lattice. Previously
the recursion method has been used to examine the effects
of disorder in $s$-wave\cite{annett,LMG} and $d$-wave\cite{xiang} 
superconductors, and to determine the core structure of vortices
and explain the origin of
de Haas van Alphen oscillations in the superconducting state\cite{miller}.
In particular Litak, Miller and Gy\"orffy\cite{LMG} have given 
a detailed description of the application of the recursion method to 
local interactions, corresponding to $s$-wave superconductors. Here we extend 
this to the case of non-local interactions necessary to obtain 
$d$-wave superconductivity and apply the method to various
interfaces of $s$ and $d$-wave superconductors.

The method of performing our self-consistent calculations will be
described in section~\ref{sec2}.
Here we introduce the Bogoliubov de Gennes equation 
with a general interaction, $U_{ij}$, and
demonstrate how this general Hamiltonian can be solved 
self-consistently using the recursion
method \cite{RHaydock,LMG}.  A necessary and non-trivial
step in the calculation, as described below,
involves finding the density of states accurately
by extrapolation of the recursion method continued fraction.

In Sec.~\ref{sec3} we proceed to apply the method to several
different problems. Firstly various test calculations
are described, including:   the local density of
states for a uniform system with no interactions,
the local density of states for a system with a local
attractive interaction (local $s$-wave superconducting order parameter)
and a system with a non-local attractive interaction. We show
that for this non-local interaction there are two possible solutions,
these solutions being a non-local $s$-wave superconducting order
parameter (extended $s$-wave) 
and a non-local $d$-wave superconducting order parameter.

Having tested the method on uniform systems we present our self-consistent
solutions for the interface between two different materials. We will
consider three different interfaces. First  we consider a
normal metal to $s$-wave superconductor,
$N-S^{s}$, interface where the pairing interaction is zero in the normal region
($N$) and purely local in the
superconducting region ($S^{s}$). Then we will consider a 
$d$-wave to normal metal,
$S^{d}-N$,
interface where again the interaction in the normal region is zero 
and there is a non-local attractive interaction in the
superconducting region ($S^{d}$). Finally a study of a
$S^{d}-S^{s}$ interface will be described. These three calculations
enable us to make a comparative study
of how the local density of states and order parameter
changes as a function of position across the
different types of interface.

\section{Theory/Model.}
\label{sec2}

\subsection{The Bogoliubov de Gennes Equation.}
\label{bdge}

The Bogoliubov de Gennes 
equation on a tight binding square lattice
has the form
\begin{equation}
\sum_{j} \mb{H_{ij}} 
\left( \begin{array}{c}
u_{j}^{n} \\
v_{j}^{n}
\end{array}
\right)
=
E_{n}
\left(
\begin{array}{c}
u_{i}^{n}\\
v_{i}^{n}
\end{array}
\right) \label{eq:1}
\end{equation}
with

\begin{equation}
\label{eq:1a}
\mb{H_{ij}}= \left( \begin{array}{cc}
H_{ij} & \Delta_{ij}  \\
\Delta^{\star}_{ij} & 
-H^{\star}_{ij}
\end{array} \right)
\end{equation}
where $u_i^n$ and $v_i^n$ are the particle and hole amplitudes, on
site $i$, associated with an eigenenergy $E_n$ and
where $\Delta_{ij}$ is the (possibly non-local)
pairing potential or gap function.

In the fully self-consistent Bogoliubov de Gennes equation the
normal state Hamiltonian $H_{ij}$ is given by
\begin{equation}
\label{eq:2}
H_{ij}=(t_{ij}+\frac{1}{2} U_{ij} n_{ij})(1-\delta_{ij}) +
(\epsilon_{i} - \mu + \frac{1}{2}U_{i} n_{i})\delta_{ij}
\end{equation}
where $\mu$ is the
chemical potential,  $\epsilon_{i}$ is the
normal on site energy of site $i$ and $t_{ij}$ is the hopping
integral between site $i$ and site $j$, for the rest of this paper
$t_{ij}$ is non zero for nearest neighbours only. The on-site
and off-site interaction terms $\frac{1}{2} U_{ii} n_{ii} $ and
$\frac{1}{2} U_{ij} n_{ij}$ are the Hartree-Fock potentials
corresponding to the on-site interaction $U_i$ and the
non-local interaction $U_{ij}$. 
The charge density entering the Hartree-Fock terms
$n_{ij}$  is given by

\begin{equation}
\label{eq:3}
n_{ij}=\sum_{\sigma}\langle\Psi_{i \sigma}^{\dagger}\Psi_{j \sigma}\rangle=
2\sum_{n}((u_i^n)^{\star}u^n_j f(E_n)+v_i^n(v_j^n)^{\star}(1-f(E_n)).
\end{equation}
Similarly the pairing potentials are defined as

\begin{equation}
\label{eq:4}
\Delta_{ij}=-U_{ij}F_{ij}
\end{equation}
where the anomalous density is

\begin{equation}
\label{eq:5}
F_{ij}=\langle\Psi_{i \uparrow}\Psi_{j \downarrow}\rangle=
\sum_{n}(u_i^n (v^n_j)^{\star}(1- f(E_n))
-(v_i^n)^{\star}(u_j^n)f(E_n).
\end{equation}
In equations (\ref{eq:3}) and (\ref{eq:5}) the sums only consider
terms $E_n$ up to the condensate chemical potential ($\mu$).

A solution to the above system of equations will be fully self-consistent
provided that both the normal ($U_{ij}n_{ij}$) and anomalous 
($\Delta_{ij}$) potentials are determined consistently
with the corresponding densities
$n_{ij}$ and $F_{ij}$ via Eqs. \ref{eq:2}  and \ref{eq:4}.
Note that the normal Hartree-Fock terms $U_{ij}n_{ij}$ play
an important role and cannot be neglected. For on-site interactions these
terms correspond to position dependent shifts in the on-site energy,
while for non-local interactions these terms renormalise the
hopping $t_{ij}$ leading to position dependent changes in
the electronic bandwidth.

Figure 1 illustrates the geometry corresponding to
this system of equations.  The tight-binding lattice 
has nearest neighbour hopping interactions ($t_{ij}$),
as well as  a coupling between particle and hole
space, via a superconducting order parameter ($\Delta_{ij}$). 
If the interactions are purely on-site ($U_i$) attractions 
then the pairing potential will be purely local ($\Delta_{ii}$),
corresponding to the dashed line in Fig. 1. On the other hand
when the interaction is non-local ($U_{ij}$, $i \neq j$)
the pairing potential $\Delta_{ij}$ will also be non-local,
as illustrated by the solid lines in Fig. 1. For computational
convenience we limit both the hopping and non-local interaction
to nearest neighbours distances.
We also
need to specify over what energy range the interaction has an effect, and
as in BCS \cite{BCS}, we will assume that it 
only  acts over a small energy range centred on the Fermi energy,
$ \mu \pm E_c$.

\subsection{The Recursion Method.}
\label{Recursion}

The method we have adopted to solve  the above system of equations
is the
recursion method \cite{RHaydock}.
This method allows us to calculate the electronic Green's functions

\begin{equation}
\label{eq:6}
G_{\alpha \, \alpha^{\prime}}(i,j,E)=
\langle i \alpha|\frac{1}{E \mb{1}- \mb{H}} |j \alpha^{\prime}\rangle
\end{equation}
where the indices $i$ and $j$ denote sites,
while   $\alpha$ and $\alpha^{\prime}$ represent the 
particle or hole
degree of
freedom on each site. We denote particle degrees of freedom by
$\alpha=+$ and hole degrees of freedom by $\alpha=-$. For example
$G_{+ \, -}(i,j,E)$ represents the Greens function between the
particle degree of freedom on site $i$ and the hole degree of freedom
on site $j$.

To compute the Green's functions (\ref{eq:6}) we can closely follow the method
described by
Litak, Miller and Gy\"orffy\cite{LMG} 
for the special case of a local interaction ($U_{ij} = U_{ii} 
\delta_{ij}$). 
Using their method we can transform the Hamiltonian to a block
tridiagonal form

\begin{equation}
E \mb{1}- \mb{H}=
\left(\begin{array}{cccccccc}
E\mb{1}-\mb{a_0} & -\mb{b_1} & 0 & 0 & 0 & 0 & 0 & \cdots \\
-\mb{b^{\dagger}_1} & E\mb{1}-\mb{a_1} & -\mb{b_2} & 0 & 0 & 0 & 0 & \cdots \\
0 & -\mb{b^{\dagger}_2} & \ddots & \ddots & 0 & 0 & 0 &\cdots \\
0 & 0 & \ddots & \ddots & \ddots & 0 & 0 & \cdots \\
0 & 0 & 0 & -\mb{b^{\dagger}_n} & E\mb{1}-\mb{a_n} & -\mb{b_{n+1}} & 0 &
\cdots \\
\vdots & \vdots & \vdots & 0  & \ddots & \ddots & \ddots & \ddots 
\end{array}
\right)
\label{eq:7}
\end{equation}
where $\mb{a_n}$ and $\mb{b_n}$ are $2 \times 2$ matrices.  Given this
form for $\langle i \alpha|E \mb{1}- \mb{H}|j \alpha^{\prime}\rangle$
and expressing the Green's function as

\begin{equation}
G_{\alpha \, \alpha^{\prime}}(i,j,E)=
\langle i \alpha|(E \mb{1}- \mb{H})^{-1} |j \alpha^{\prime}\rangle,
\label{eq:8}
\end{equation}
the Greens functions above can be evaluated as a matrix continued
fraction so that

\begin{equation} 
\mb{G}(i,j,E)= 
\left(E\mb{1}-\mb{a_{0}}-\mb{b^{\dagger}_1} 
\left(E\mb{1}-\mb{a_{1}}-\mb{b^{\dagger}_2} 
\left(E\mb{1}-\mb{a_{2}}-\mb{b^{\dagger}_3} 
\left( E\mb{1}-\mb{a_{3}}-\ldots \right) ^{-1} \mb{b_3} 
\right) ^{-1}\mb{b_2} \right) ^{-1}\mb{b_1} \right) ^{-1} 
\label{eq:9}
\end{equation}
where

\begin{equation}
\mb{G}(i,j,E)=\left(
\begin{array}{cc}
G_{\alpha \, \alpha}(i,i,E) & G_{\alpha \, \alpha^{\prime}}(i,j,E) \\
G_{\alpha^{\prime} \, \alpha}(j,i,E) & 
G_{\alpha^{\prime} \, \alpha^{\prime}}(j,j,E)
\end{array}
\right).
\label{eq:10}
\end{equation}

Within equations (\ref{eq:7}) and (\ref{eq:9}) we have a formally exact
representation of the Green's functions.  However in general 
both the tridiagonal
representation of the Hamiltonian, and the 
matrix continued fraction 
(\ref{eq:7}) will be infinite.  In practice one can only calculate
a finite number of terms in the continued fraction exactly.
In the terminology of the recursion method
it is necessary to {\em terminate} the continued fraction
\cite{RHaydock,LMG,AMagnus,DTT,CMMNex,GAllan,TKW}.

If we were to calculate up to
and including $\mb{a_n}$ and $\mb{b_{n}}$ and then simply set subsequent
coefficients to zero
then the Green's function
would have $2n$ poles along the real axis.  The density of states
would then correspond to a set of $2n$ delta functions.
Integrated quantities such as the the densities
$n_{ij}$ and $F_{ij}$ could depend strongly on $n$, especially
since only a few of the $2n$ delta functions would be within the relevant
energy range within the  BCS cut off, $E_c$. In order to 
obtain accurate results it would be necessary to calculate a large number
of exact levels, which would be expensive in terms of both computer
time and memory.

As a more efficient alternative we choose to terminate the continued
fraction using the extrapolation method, as used previously
by Litak, Miller and Gy\"orffy\cite{LMG}.
We calculate the values for
$\mb{a_n}$ and $\mb{b_{n}}$ exactly up to the first $m$ coefficients
using the recursion method.  Then, noting the fact that the elements of
the matrices
$\mb{a_n}$ and $\mb{b_{n}}$ vary in a predictable manner \cite{LMG}, we
extrapolate the elements of the matrices for a further $k$ iterations,
where $k$ is usually very much greater than $m$.  This enables
us to compute the various densities of states, and the
charge densities $n_{ij}$ and $F_{ij}$ accurately
with relatively little computer time and memory.

In terms of the Green's functions $G_{\alpha \,
\alpha^{\prime}}(i,j,E)$ the pairing and normal Hartree-Fock
potentials  
$\Delta_{ij}$ and $\frac{1}{2}U_{ij}n_{ij}$ 
are given by

\begin{equation}
\label{eq:12}
\Delta_{ij}=\frac{1}{2 \pi} U_{ij}
\int_{-E_c}^{E_c} (G_{+ \, -}(i,j,E+\imath \eta) - 
G_{+ \, -}(i,j,E-\imath \eta))(1-f(E))dE
\end{equation}
and

\begin{equation}
\frac{1}{2}U_{ij}n_{ij}=\frac{1}{2\pi} U_{ij}
\int_{-E_c}^{E_c} (G_{+ \, +}(i,j,E+\imath \eta) - 
G_{+ \, +}(i,j,E-\imath \eta))f(E)dE.
\label{eq:13}
\end{equation}
where $\eta$ is a small positive number.
 
To obtain the above equations we have used the fact that

\begin{equation}
\left(
\begin{array}{c}
u_i^n \\
v^n_i
\end{array}
\right)
\, \, \, \, {\rm and} \, \, \, \,
\left(
\begin{array}{c}
-(v_i^n)^{\star} \\
(u^n_i)^{\star}
\end{array}
\right)
\end{equation}
are the eigenvectors of equation (\ref{eq:1}) with eigenvalues $E_n$
and $-E_n$ has been used. Also note that the integrals in equations
(\ref{eq:12}) and (\ref{eq:13}) are bounded by the cut-off $E_c$,
corresponding to the energy dependent interaction

\begin{equation}
U_{ij}(E)=
\left\{
\begin{array}{c}
-|U| \, \, {\rm for} \, \, |E-\mu| \le E_c\\
0  \, \, {\rm for} \, \, |E-\mu| > E_c
\end{array}
\right. ,
\end{equation}
as in BCS theory. Our cut-off $E_c$ can  correspond to the BCS cut-off
$\hbar \omega_D$ arising from retardation of the electron-phonon
interaction, or any other energy scale cut-off for the interaction which may be applicable for high temperature superconductors

\subsection{Achieving Self-consistency.}

Using the above methods to calculate $\Delta_{ij}$ and $U_{ij}n_{ij}$ 
we need to achieve a fully self-consistent solution.
Firstly we make use of any symmetries in the system
in order to minimise the number of calculations which are necessary.
For example on an infinite square lattice with no variation in
any of the potentials only one independent site needs to be calculated 
since this
site can be mapped onto all of the other sites.  Secondly,
once we have decided
which sites need to be calculated self-consistently $\Delta_{ij}$ and 
$n_{ij}$ can be calculated for those sites, remembering that
on a square lattice each site will have four nearest neighbours.
This implies that in general
 we will have to calculate equation nine different Green's functions
in order to calculate $\Delta_{ij}$ and $n_{ij}$. This can be seen by
considering site $i$ in figure 1 and noting that we need to calculate
the Greens functions shown in table I, depending on whether the interaction is
purely local, purely non-local, or both local and non-local.
Having calculated the appropriate Green's functions new
values for $\Delta_{ij}$ and $n_{ij}$ can be calculated, which we 
will denote as
$\Delta^{(1)}_{ij}$ and $n^{(1)}_{ij}$.
Inserting these into the Hamiltonian and repeating the calculation of then
Green's functions leads to a new set
$\Delta^{(2)}_{ij}$ and $n^{(2)}_{ij}$  and so on.
 We repeat this iteration for all $i$ and $j$ until

\begin{equation}
\label{eq:14}
\left|\frac{|\Delta^{(n-1)}|-|\Delta^{(n)}|}{|\Delta^{(n)}|}\right| \le 0.001
\end{equation}
and 

\begin{equation}
\label{eq:15}
\left|\frac{|n^{(n-1)}|-|n^{(n)}|}{|n^{(n)}|}\right| \le 0.001.
\end{equation}

Since $\Delta_{ij}$ and $n_{ij}$ can be complex we need to also check
for convergence in their associated phases. We do this and find that
convergence in the phase gradient  of the complex parameters is much more rapid
than the convergence in the magnitude.

\section{Numerical Results.}
\label{sec3}

\subsection{Uniform Systems.}
\label{us}

As a first test of the above methods let us examine a
bulk superconductor, corresponding to an infinite 2-d square
lattice with either local or non-local attraction.
These examples will show how well
such quantities as the local particle density of states can be
calculated and how the extrapolation of the elements of the matrices
$\mb{a_n}$ and $\mb{b_{n}}$ is performed.

The quantity of  interest is the local particle density of
states, which can be calculated from the following expression

\begin{equation}
\label{eq:16}
N_i(E)=\frac{1}{2\pi} (G_{+ \, +}(i,i,E+\imath \eta)-G_{+ \,
+}(i,i,E-\imath \eta)).
\end{equation}

Consider  first a
non-interacting system where $U_{ij}=0$, $\mu=0$, $\epsilon_{i}=0$,
and  $t_{ij}=1$
for nearest neighbours and zero everywhere else. Figure 2 
shows the local particle density of states for a calculation
where the number of exact continued fraction
levels was $m=50$ and the elements of $\mb{a_n}$ and $\mb{b_{n}}$ were
extrapolated for $2000$ more values.  For this calculation
the convergence parameter
$\eta$ was chosen as
$\eta=0.02$.  Fig. 2 shows that using the extrapolation method the 
central logarithmic van 
Hove singularity and the sharp band edges
can be resolved very well. Figure 3 shows 
the first $100$ continued fraction coefficients 
 $\Re b_n^{1 1}$, where the first
$50$ are calculated directly using the recursion method and the rest
are the extrapolated values. It is clear from the figure that
the oscillations in $\Re b_n^{1 1}$  still
persist after the first $50$ continued fraction levels. In fact
these oscillations die off slowly as $1/n$ 
and it is critical to include them correctly. 
From figure 3 it is clear that the first 50 levels provide enough 
information about the decaying oscillation that the $\Re b_n^{1 1}$  can be
extrapolated quite easily.

Having considered a system where the interaction is zero the next step
is to consider systems where the interaction is uniform and
finite. For such systems the local particle density of states can be
calculated, for different types of interaction. In figure 4 we have
plotted two different local particle densities of states for
a local interaction $U_{ij} =-2.5 \delta_{ij}$ (dashed line) and
$U_{ij} =-2.5 (1-\delta_{ij})$ for nearest neighbours 
(solid line). In each case $E_c=4$ and
$t_{ij}=1$ (for nearest neighbours) and all other parameters were set
to zero throughout the lattice. 

The dashed line in figure 4 clearly shows the energy gap at the Fermi
energy, characteristic of $s$-wave superconductivity.  The van Hove
peak in the density of states is also very clearly resolved.
The 
solid line in Fig. 4
shows the local particle density of states going to zero at the Fermi
energy, in a manner which is typical of the local particle density of
states for a $d$-wave superconductor. In this case of the non-local
interaction the order parameter changes sign we as rotate by $\pi/2$
around a site, ie. in reference to figure 1
$\Delta_{ij_1}=-\Delta_{ij_2}$, $\Delta_{ij_2}=-\Delta_{ij_3}$ and
$\Delta_{ij_4}=-\Delta_{ij_3}$. 
The way we have performed the
calculation is to keep the Fermi energies the same in the two
calculations but change, in the case of the local interaction
(dashed line), the density and, in the case of the non-local
interaction (solid line), the width of the band self-consistently. This
has the effect of moving the
system away from half filling in the case of the local interaction, 
and, broadening the band in the case of the non-local
interaction, because of  
the local and non-local Hartree-Fock terms in the Hamiltonian 
$U_{ii}n_{ii}$ and $U_{ij}n_{ij}$ respectively.

At this point one should note that in the case of a non-local
interaction one can as well as having a $d$-wave self-consistent
solution to the Bogoliubov de Gennes equation, it is also possible to
obtain an extended $s$-wave solution, ie. $\Delta_{ij_1}=\Delta_{ij_2}$,
$\Delta_{ij_2}=\Delta_{ij_3}$ and $\Delta_{ij_4}=\Delta_{ij_3}$.
However we find
that such solutions are less stable than the $d$-wave solutions, this is only true at or near half filling of the band.

To obtain the results shown in figure 4 we have again calculated $50$ levels
of the recursion method exactly and then extrapolated for further
$2000$ levels. This can easily be done because the elements of
$\mb{a_n}$ and $\mb{b_{n}}$ vary in a predictable manner, as has
already been seen for the case without interactions.

\subsection{Interfaces.}
\label{int}

Having considered systems where the interactions remain uniform
throughout the structure the next step is to consider systems which
contain interaction strengths which vary in real space. The most
simple case one can conceive for this scenario is an interface. We
will simply model the interface by allowing the interaction
to change in a step like manner. 

We will consider three separate situations, $N-S^{s}$, $S^{d}-N$ and
$S^{d}-S^{s}$. In the normal region we shall
set $U_{ij}=0$, hence the order parameter in this region will be zero,
(but one should note that this does not imply that $F_{ij}$ is
zero). Before we look at the numerical results it is worthwhile
considering what one may expect to find. In the case of a local
interaction the results are well documented, i.e. the magnitude of the
superconducting order parameter reaches a maximum at the bulk value
a few coherence
lengths in the superconducting region away from the normal
interface. In the case of the non-local interaction we would also
expect the amplitude of the superconducting order parameter to reach a
maximum several coherence lengths away from the interface, but the
problem of how to define the magnitude of the superconducting order
parameter now arises. Going back to figure 1 we can see that for each
site $i$ there are five $\Delta_{ij}$'s, so hence for each site we can
define five order parameters per site. We can also combine these
different order parameters on each site in the following manner

\begin{equation}
|\Delta^{(s(local))}_i|=|\Delta_i|
\end{equation}

\begin{equation}
|\Delta^{(d)}_i|=\frac{1}{4}|\Delta_{ij_{1}}-\Delta_{ij_{2}}+\Delta_{ij_{3}}-
\Delta_{ij_{4}}|
\end{equation}

\begin{equation}
|\Delta^{(s(non-local))}_i|=\frac{1}{4}|\Delta_{ij_{1}}+\Delta_{ij_{2}}+
\Delta_{ij_{3}}+\Delta_{ij_{4}}|
\end{equation}
so that each equation defines a different type of symmetry for that
site. Since the systems we are interested in change in the
$x$-direction only it is possible, when one is considering the
properties of that interface, to look along one line of sites in the
$x$-direction and note that for any other $y$ coordinate the properties
of the system are the same, so $\Delta_i \rightarrow \Delta(x)$.

Having defined all the quantities of interest the next step is to
specify some of the systems of interest. The three systems we are
going to consider are as already pointed out $N-S^{s}$, $S^{d}-N$ and
$S^{d}-S^{s}$, to set up these systems we used the parameters shown in
table II.  

Figures 5(a-c) plot the three main symmetry components of the
order parameter 
 $|\Delta^{(s(local))}(x)|$ (dashed
line), $|\Delta^{(d)}(x)|$ (circles) and
$|\Delta^{(s(non-local))}(x)|$ (solid line) for the three
different geometries $N-S^{s}$ (figure
5(a)), $S^{d}-N$ (figure 5(b)) and $S^{d}-S^{s}$ (figure 5(c)). 
The interface corresponds to $x=100$ on the figures.
Figure
5(a) shows, as expected, that the $s$-wave order parameter,
$|\Delta^{(s(local))}(x)|$, simply rises over a
coherence length to a maximum at the  bulk superconducting order parameter.
Because the interaction is purely on-site in Fig. 5(a)
$|\Delta^{(d)}(x)|=|\Delta^{(s(non-local))}(x)|=0$. 

In figure 5(b)
we see that for the $d$-wave to normal metal interface
$|\Delta^{(d)}(x)|$ also drops to zero at the interface.
However, unlike the $s$-wave case, it does not simply drop to zero smoothly
but has a sharp peak structure right at the interface. 
The origin of this peak is explained by looking at the extended $s$-wave
component, $|\Delta^{(s(non-local))}(x)|$ (solid line in Fig. 5(b)). 
 We see that the extended $s$-wave gap function is
finite near the interface. This is due to he fact that the order
parameter varies near the interface and hence $\Delta_{j_{1}}(x) \ne
\Delta_{j_{3}}(x)$, making the values of
$|\Delta^{(s(non-local))}(x)|$. This is emphasised in figure 5(d)
where $\Delta_{j_3}(x)-\Delta_{j_1}(x)$ is plotted, from this graph one
can see that the peak in $|\Delta^{(d)}(x)|$, in figure 5(b), near the
interface is due to the component in the the $x$ direction. 

Figure
5(c) shows the $d$-wave to $s$-wave superconductor interface.
Again we can see that the extended $s$-wave component 
$|\Delta^{(s(non-local))}(x)|$ is non-zero at the interface, even though
 it is zero in the bulk on both sides, and that this leads to sharp 
features in both the local $s$-wave and $d$-wave order parameters
near the interface.

Having seen how the profiles of the superconducting order parameters
are affected by the proximity of different materials, we now look
at how the local particle density of states changes as we move across
the various interfaces.  Figures 6, 7 and 8 are contour plots of the
local particle densities of states for the three interfaces of
interest. Figure 6 shows a contour plot for the $N-S^{s}$
interface. Looking at this plot one can see that as we move across the
interface, at $x=100$, the superconducting gap opens up
within a couple of atomic sites. On the normal metal side,
for $x<100$,
the van Hove singularity in the centre of the band can be clearly
seen, but as we move into the superconducting region the band edges
are shifted (due to the Hartree-Fock potential  term)
and the superconducting gap opens up at $E=0$. In the
superconducting region the van-Hove singularity is shifted away from
$E=0$ as can also be seen in figure 3 (dashed line). Due to the mismatch in
the band edges we see oscillations in the local particle
density of states near the band edges; these are simply Friedel oscillations.

Figure 7 shows a similar contour plot of the local particle density of states
for an $S^{d}-N$ interface.  Again we can clearly see the gap
in the superconducting region and the van Hove singularity
in the normal region.  In this system the Hartree-Fock potential term
leads to an increase in overall band width on the d-wave side. Again
since the band edges do not match up
we see Friedel oscillations in the local particle density of states near the
band edges.

Finally in figure 8 we have plotted the local particle density of states as we
move across the $S^d-S^s$ interface. This plot has many interesting
features, the first to note is that again due to the mismatch in the
band edges oscillations appear in the local particle density of
states. Secondly for $x <100$ ($S_{d}$ region) the density of states
gradually goes to zero at $E=0$ (typical of d-wave superconductivity
(see figure 4 (solid line))). Whereas for the $S^s$ region the local
particle density of states drops to zero very sharply. The main points
of interest is what happens at the interface itself. In plane of
the interface there are states in the gap, as both the d-wave and
s-wave order parameters are suppressed. At $x=100$ there are two peaks
in the density of states
just above and below $E=0$, which as we move further into the $S^s$
region are shifted to become the BCS density of
states singularities just above and below the superconducting
gap.  Note that the parameters for the calculation
in Fig. 8 were chosen so that $|\Delta^{(d)}| >> \Delta^{(s(local))}|$
as would be the case for a YBCO-Pb junction such as those
used by Wollman {\it et al.} \cite{wollman}.

\section{Conclusions.}

In this paper we have shown how it is possible to perform
self-consistent calculations of the Bogoliubov de Gennes equation,
using the recursion method. This method has the advantage of being an
order $N$ method and hence allows us to tackle problems with a
relatively small amount of computational effort.  A key to obtaining accurate
densities of states with relatively little computational effort is the
extrapolation procedure we have used to terminate the matrix continued
fraction. Our method is fully self-consistent, including both self-consistency
in the order parameter and in the normal Hartree-Fock potentials.
As we have shown these normal potentials make significant contributions
by shifting or widening the density of states in a spatially
dependent manner.  Our method can deal with both local
attractive interactions, corresponding to local $s$-wave superconductivity,
or non-local interactions 
corresponding to $d$-wave or extended $s$-wave pairing. In our system
we found that the $d$-wave is more
stable. 

As a first application of the method we examined three simple
interfaces, corresponding to an $s$-wave $S-N$ junction, a $d$-wave
$S-N$ junction and an $s$-wave to $d$-wave $S-S$ junction.  The numerical
results show a number of interesting features, including a
non-monotonic variation of the order parameters near the interface, a
surface layer of extended $s$-wave pairing (even though it is not stable
in the bulk), and subtle effects of the self-consistent Hatree-Fock
terms in the Bogoliubov de Gennes Hamiltonian leading to Friedel
oscillations and spatially dependent shifts in the van Hove
singularities near the interfaces, as highlighted by the contour plot
in figure 8.

In future we hope to apply our method to more complex 
interfacial phenomena in superconductors, such as junctions 
carrying supercurrent (e.g to look for $\pi$-junctions),
superconducting twin boundaries and grain boundary
junctions.  Our methods can also be applied to many other
problems in superconductivity
such as the structure of vortex cores in $s$ or  $d$-wave 
superconductors, the effects of impurities and so on.

\section{Acknowledgments.}

This work was supported by the EPSRC under grant number GR/L22454.
We would like to thank M Leadbeater for bringing references 
\cite{MS1,MS2,MS3} to our notice and B.L Gy\"orffy and P. Miller 
for useful discussions.

\begin{figure}
\caption{This is a schematic diagram of a tight binding lattice, with particle and hole degrees of freedom, $\Delta_{ij}$ couples particles on site $i$ to holes on site $j$. The difference between local and non-local pairing is highlighted by the dashed (local pairing) and solid (non-local pairing) lines.}
\end{figure}

\begin{figure}
\caption{ The local particle density of states for a $2$-D tight binding lattice with no interactions ($U_{ij}=0$). For this system $t_{ij}=1$ for nearest neighbours only, $\mu=0$ and $\epsilon_i=0$.}
\end{figure}

\begin{figure}
\caption{A plot of the real part of $b(n)^{11}$ for the same system which was used to calculate the local particle density of states in figure 2.}
\end{figure}

\begin{figure} 
\caption{Two plots of the local particle density of states for, a local interaction ($U_{ij}\delta_{ij}=-2.5$, $U_{ij}(1-\delta_{ij})=0$ (dashed line)) and a non-local interaction ($U_{ij}\delta_{ij}=0$, $U_{ij}(1-\delta_{ij})=-2.5$ (solid line)), all the other parameters are equal to those used to obtain figure 2.}
\end{figure}

\begin{figure}
\caption{Figures 5(a-c) plot the profiles of different symmetries of the superconducting order parameter ($|\Delta^{(s(non-local))}(x)|$ (solid line), $|\Delta^{(s(local))}(x)|$ (dashed line) and $|\Delta^{(d)}(x)|$ (circles)) for different interfaces. Figures 5(a), (b) and (c) are for $N-S^{s}$, $S^{d}-N$ and $S^{d}-S^{s}$ interfaces respectively. Figure 5(d) plots $\Delta_{j_1}(x)-\Delta_{j_3}(x)$ for the $N-S^{d}$ interface. The parameters used to obtain the figures are given in table II.}
\end{figure}

\begin{figure}
\caption{This is a contour plot of the local particle density of states as a function of position, as one move across the $N-S^{s}$ interface. The steps in the contour plot are in units of $0.4$, ie. white represents $N(E)<0.4$ and black represents $N(E) > 0.32$. The parameters used to obtain this graph are given in table II.}
\end{figure}

\begin{figure}
\caption{This is a contour plot of the local particle density of states as a function of position, as one move across the $S^{d}-N$ interface. The steps in the contour plot are in units of $0.4$, ie. white represents $N(E)<0.4$ and black represents $N(E) > 0.32$. The parameters used to obtain this graph are given in table II.}
\end{figure}

\begin{figure}
\caption{This is a contour plot of the local particle density of states as a function of position, as one move across the $S^{d}-S^{s}$ interface.The steps in the contour plot are in units of $0.4$, ie. white represents $N(E)<0.4$ and black represents $N(E) > 0.32$. The parameters used to obtain this graph are given in table II.}
\end{figure}

\begin{table}
\begin{tabular}{|c||c|c|c|c|c|}
\hline
Interaction Type & 
$\mb{G_{+ -}}(i, i,E )$ & 
$\mb{G_{+ \pm}}(i, j_1,E)$  & 
$\mb{G_{+\pm}} (i, j_2,E)$ & 
$\mb{G_{+\pm}}(i, j_3,E)$ & 
$\mb{G_{+\pm}}(i, j_4,E)$ \\
\hline
$U_{ii}$ & Y & N & N & N & N   \\
\hline
$U_{ij}(1-\delta_{ij})$ & N & Y & Y & Y & Y  \\
\hline
$U_{ii}+U_{ij}(1-\delta_{ij})$ & Y & Y & Y & Y & Y \\
\hline
\end{tabular}
\caption{This table shows which Greens functions need to be calculated 
for systems with interactions which are
local, $U_{ii}$, non-local, $U_{ij}(1-\delta-{ij})$,
or both.  The site labels correspond to the notation of Fig. 1.}
\end{table}


\begin{table}
\begin{center}
\begin{tabular}{|c||c|c|c||c|c|c|}\hline
 & \multicolumn{3}{c||}{$x<100$} & \multicolumn{3}{c|}{$x \ge 100$} \\
\hline
System & $U_{ii}$ & $U_{ij}(1-\delta_{ij})$ & $E_c$ 
& $U_{ii}$ & $U_{ij}(1-\delta_{ij})$ & $E_c$  \\
\hline
$N-S^{s}$ & $0$ & $0$ & $0$  & $-2.5$ & $0$ & $4.0$  \\
\hline
$S^{d}-N$ & $0$ & $-3.5$ & $4.0$  & $0$ & $0$ & $0$  \\
\hline
$S^{d}-S^{s}$ & $0$ & $-3.5$ & $4.0$ & $-2.5$ & $0$ & $4.0$ \\
\hline
\end{tabular}
\caption{This table defines how the interactions vary in real space for 
the three different interfaces. All energies are given in units
where the nearest neighbour hopping $t_{ij}=1$.
Also $\mu=0$ everywhere and $T=0.01$.}
\end{center}
\end{table}
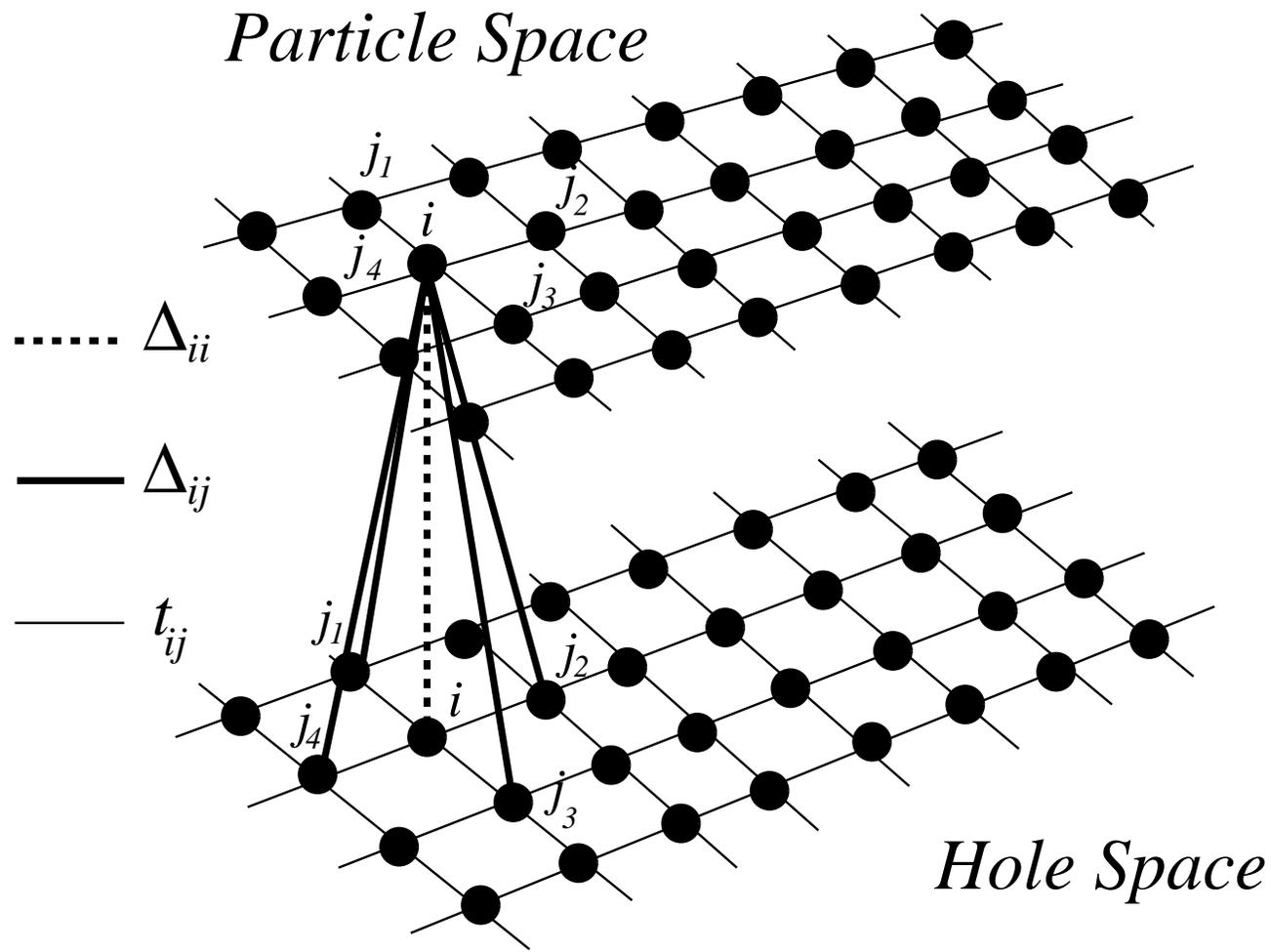
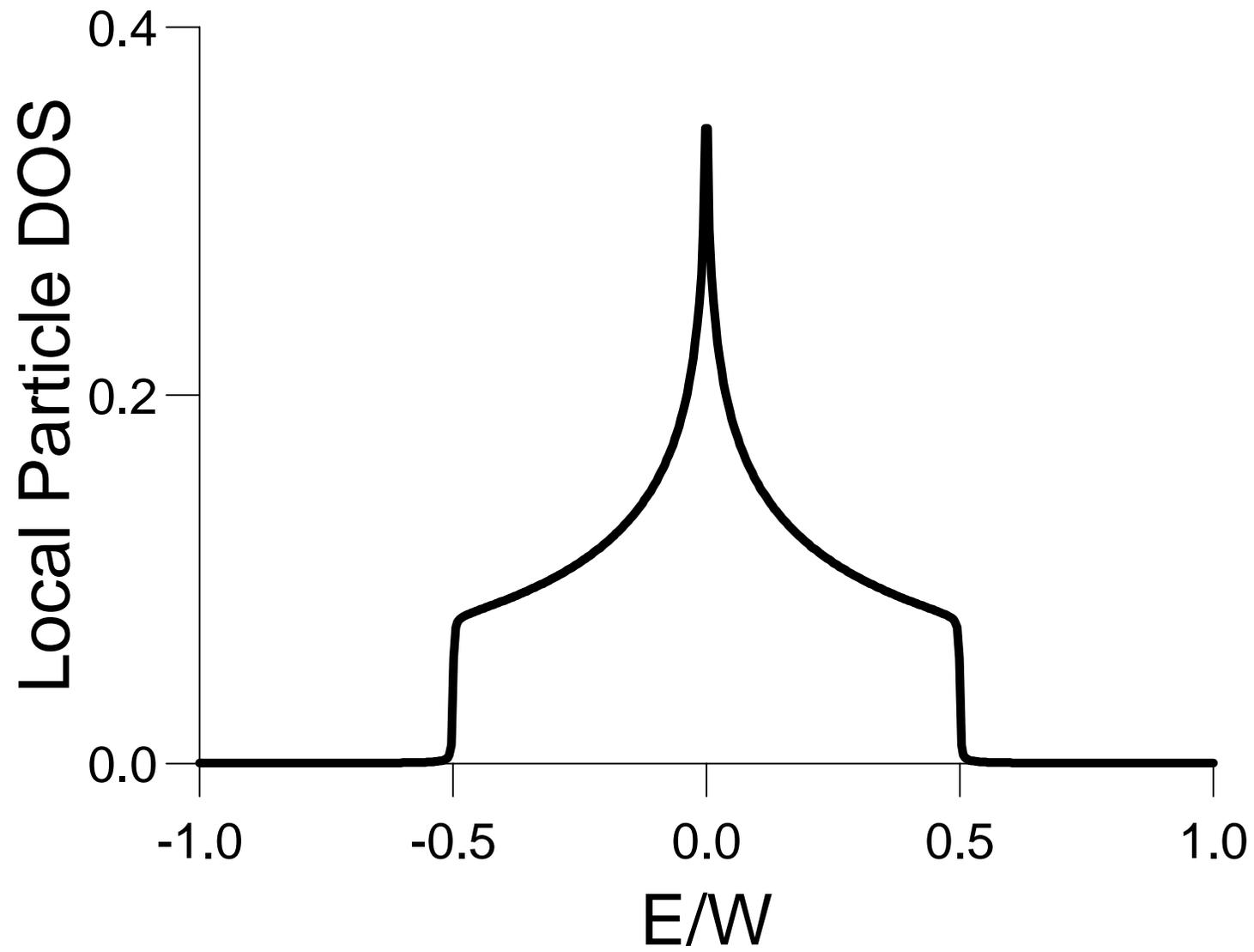
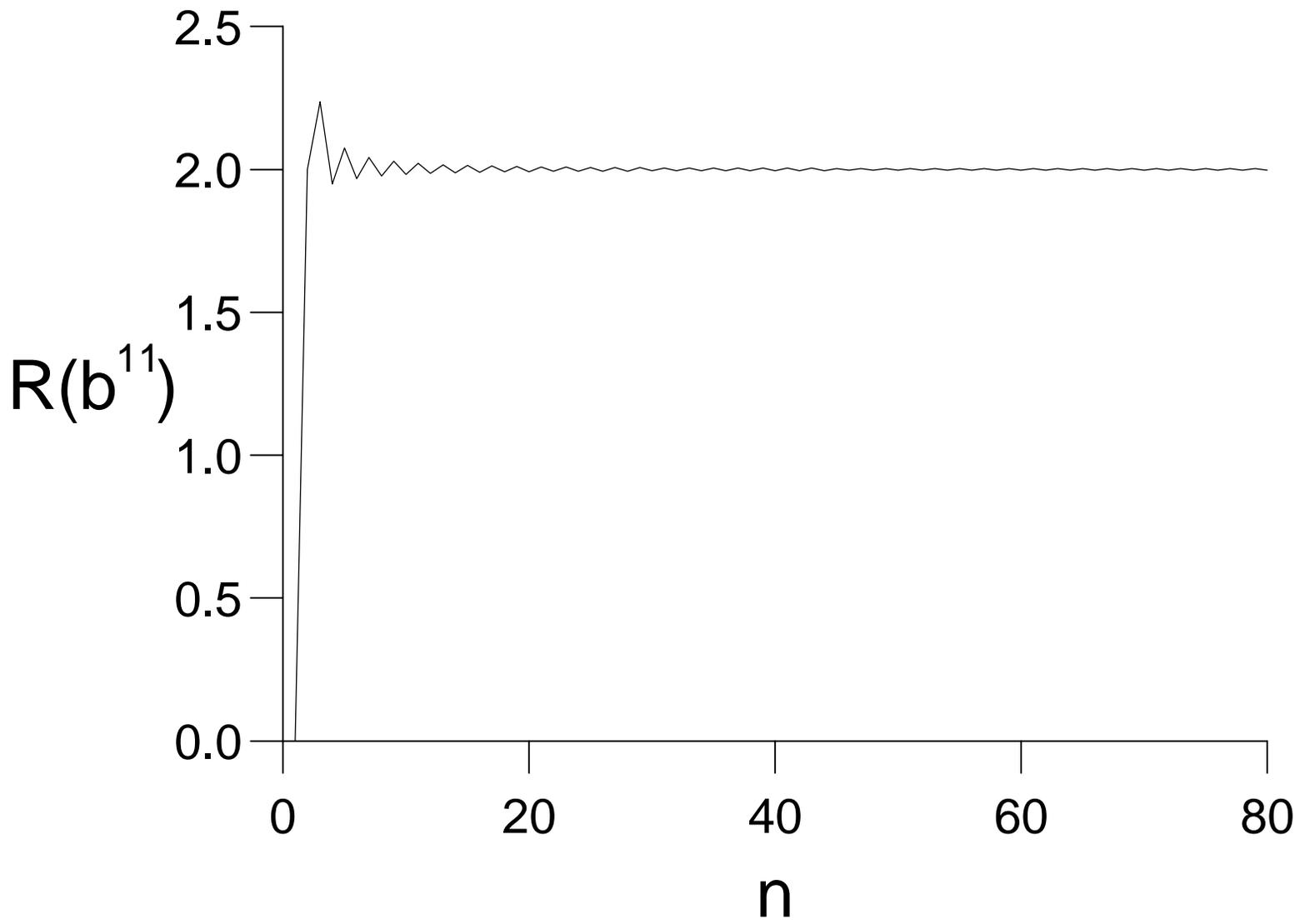
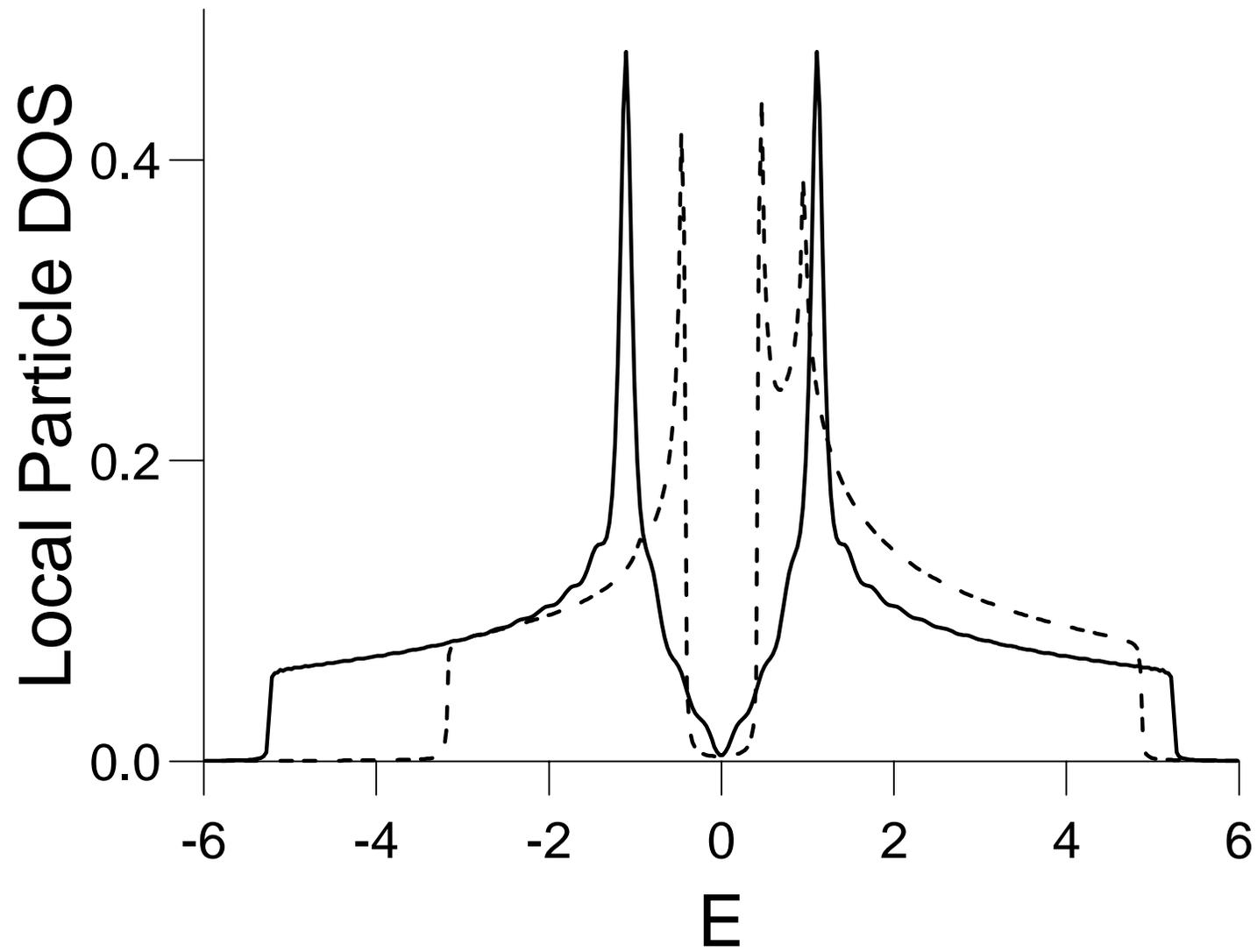
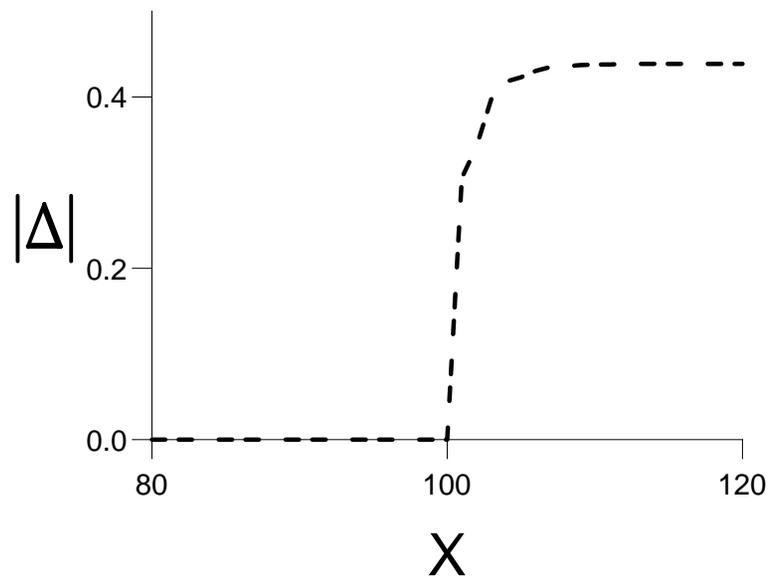
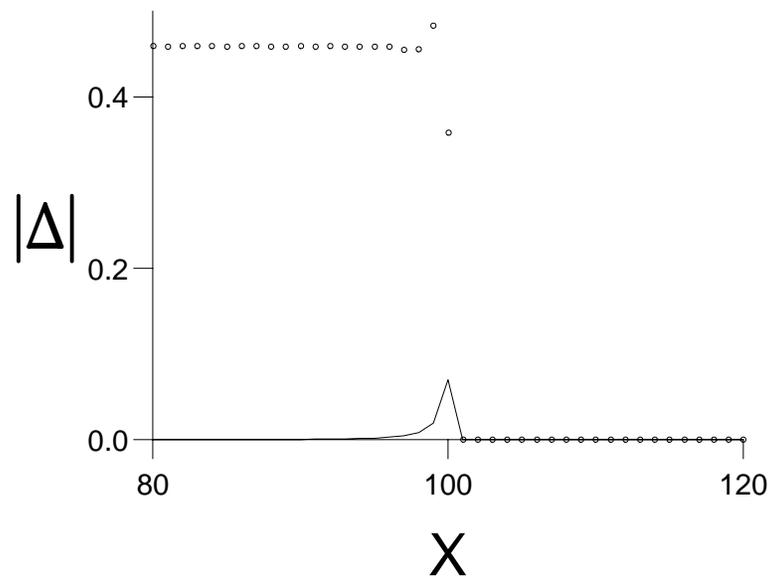
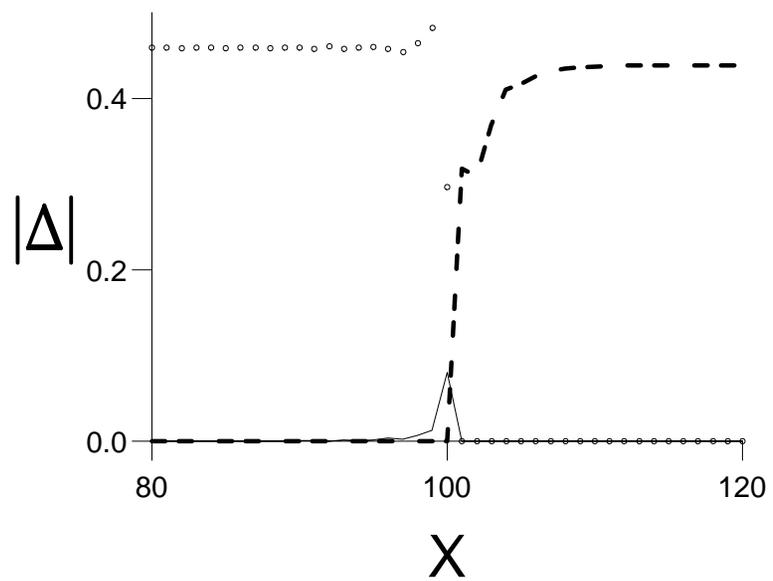
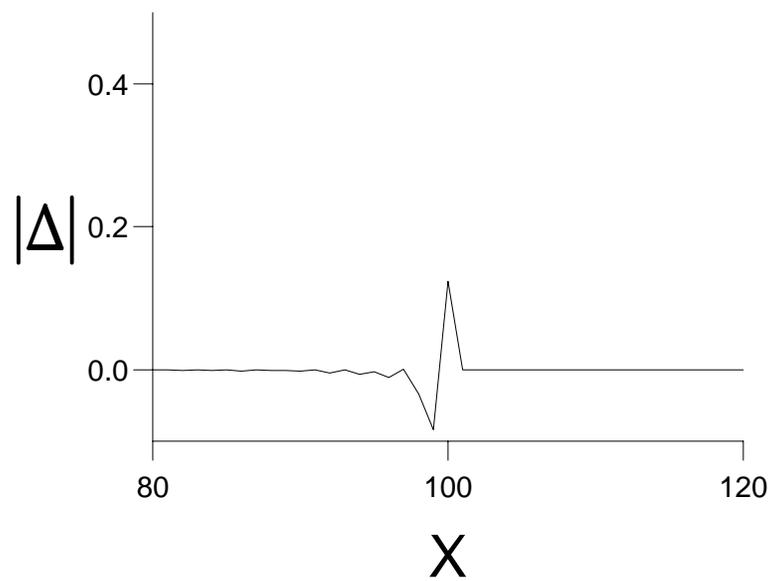
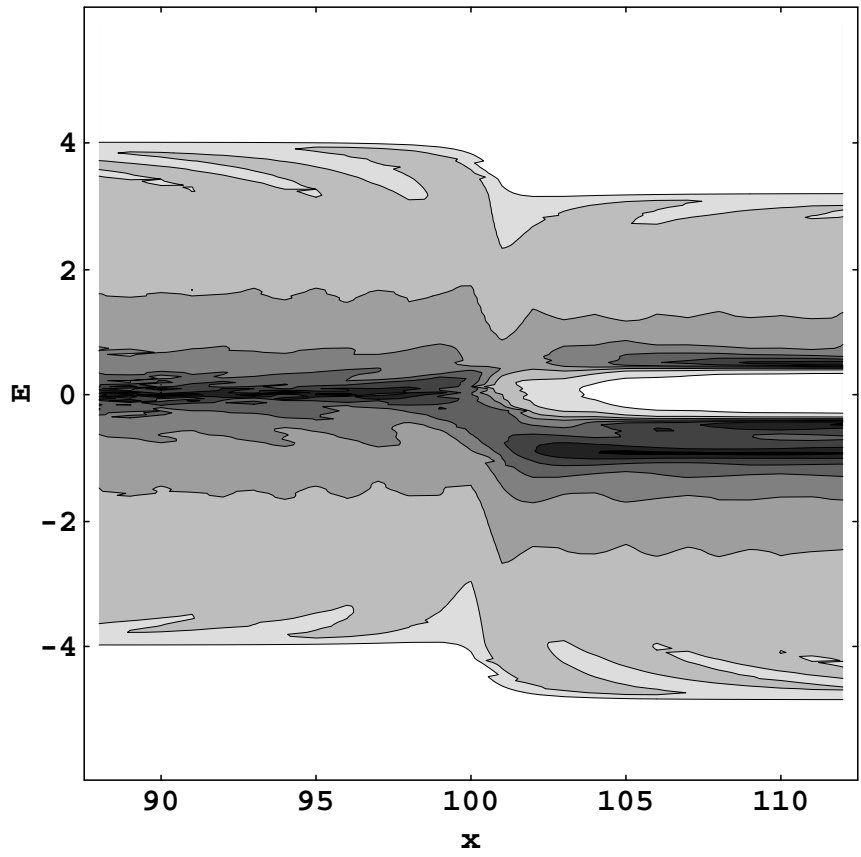
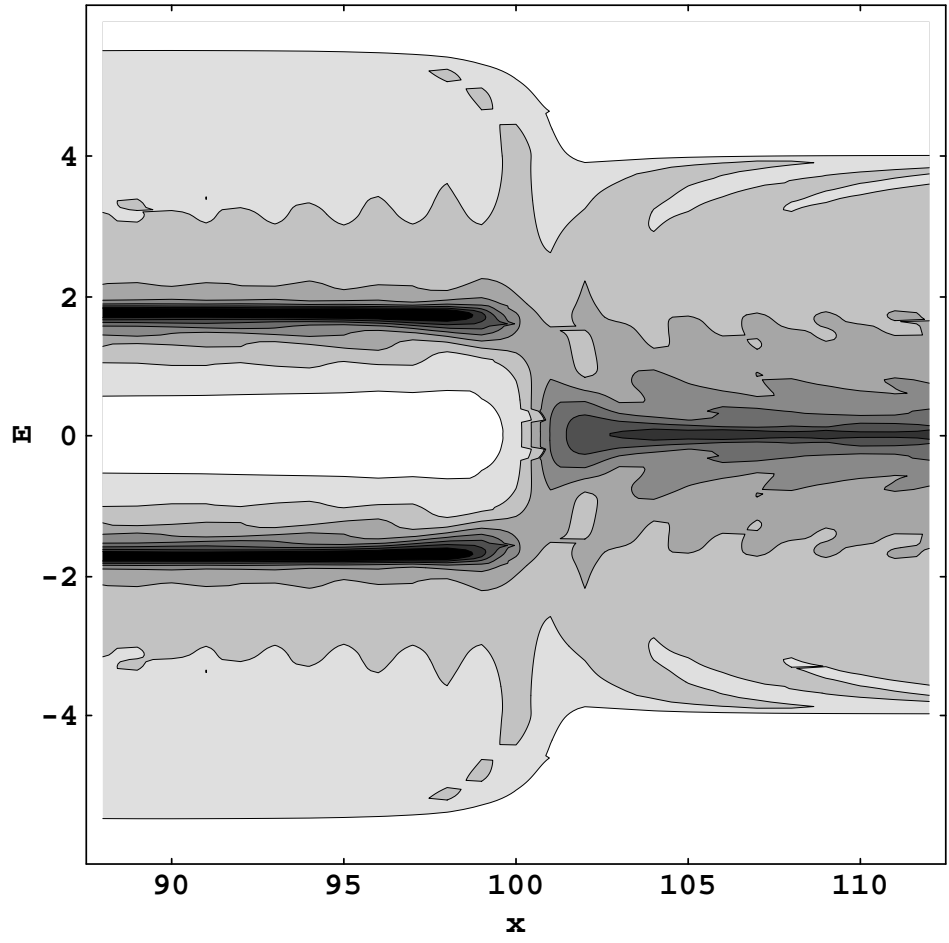
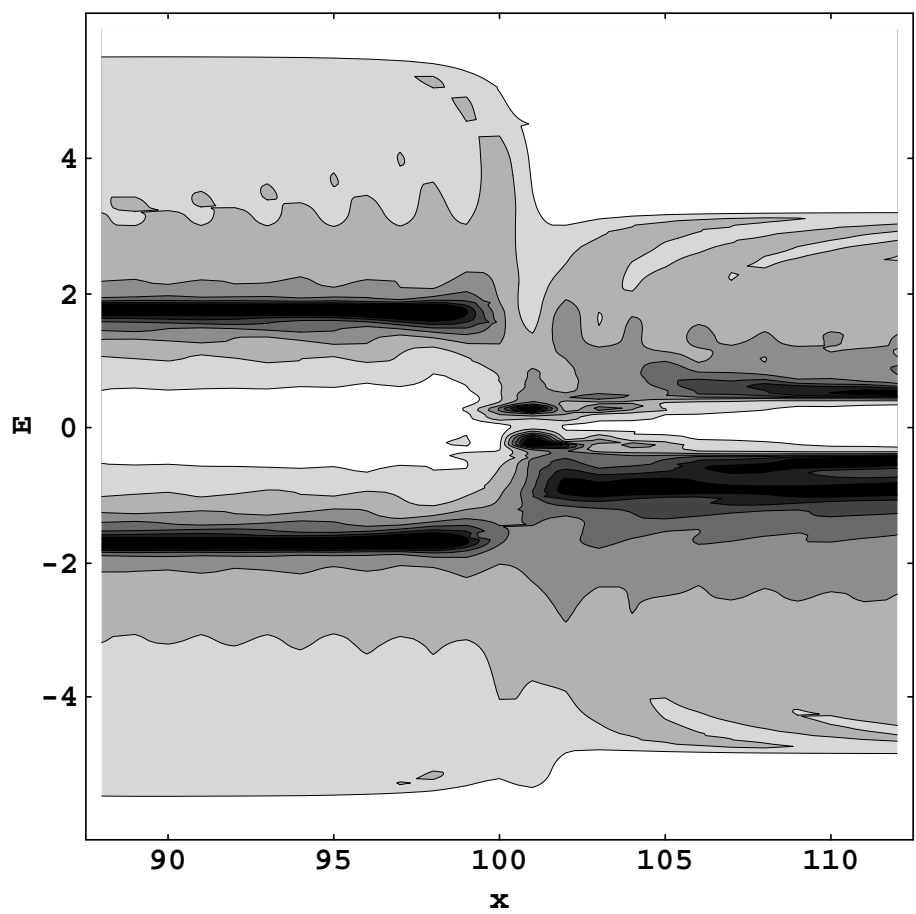
\end{document}